# Evidence for graphene plasmons in the visible spectral range probed by molecules

*Philipp Lange, Günter Kewes, Nikolai Severin, Oliver Benson, Jürgen P. Rabe*

Humboldt-Universität zu Berlin, Department of Physics, 12489 Berlin, Germany

Graphene is considered to be plasmon active only up to the infrared based on combined tight binding model and random phase approximation calculations. Here we show that the optical properties of graphene as measured by ellipsometry and simulated by density functional theory imply the existence of strongly localized graphene plasmons in the visible with a line width of 0.1 eV. Using small emitters that provide the high wavevectors necessary to excite graphene plasmons at optical frequencies we demonstrate graphene plasmon induced excitation enhancement by nearly 3 orders of magnitude.

Surface plasmons (SPs) are propagating charge density oscillations at the interface of a conductor and a dielectric. This and their strong confinement, the origin of plasmon induced field enhancement, renders them interesting for application such as light harvesting [1], Raman spectroscopy [2] and quantum information processing [3].

SPs are not limited to the interface between a conducting bulk substrate and a dielectric but they may also exist near quasi 2D conducting layers like graphene. The solution of Maxwell's equations for a p-polarized wave localized near a 2D layer with a conductivity $\sigma$ provides the plasmon dispersion relation [3,4]:

$$\lambda_{sp} \approx \frac{2\pi}{\text{Re}(i(\varepsilon+1)\varepsilon_0 \omega/\sigma)}, \qquad (1)$$

with $\lambda_{sp}$ being the wavelength and $\omega$ the frequency of the plasmon, $\varepsilon$ the dielectric function of the dielectric medium at the interface to the 2D layer and $\varepsilon_0$ the vacuum permittivity.

The existence of such a 2D plasmon with a finite wavevector requires a non-vanishing imaginary part of the conductivity $\sigma$ at the frequency of interest [3], as demonstrated for graphene in the infrared [5]. In the visible spectral range the conductivity of graphene, even for strong dopings, was, however, predicted to be primarily real, both within the tight binding model (TBM) [6] and the TBM based random phase approximation (RPA) [3]. These are the predominantly used models in the field of graphene plasmonics and graphene is thus believed to be plasmon inactive in the visible spectral range.[3,5,7] In the following we will discuss that TBM [6] and TBM based RPA [3] at optical frequencies contradicts both recent experiments and more accurate theoretical descriptions of graphene.

Precise ellipsometry measurements of the graphene refractive index for the visible spectral range have been recently reported.[8,9] The refractive index $n$ of graphene can be converted into its sheet conductivity $\sigma$ using the relation [10-12]:

$$\sigma = i\omega\varepsilon_0 d(n^2 - 1), \qquad (2)$$

with $d$ being the thickness of the graphene layer. Substitution of the *experimentally acquired* refractive index of graphene into equation 2 reveals that the real and imaginary parts of graphene's complex conductivity in the visible range are of comparable magnitude (at least for graphene on a solid substrate), in contrast to the theoretical prediction [3,6].

The real part of the graphene conductivity derived from the refractive index (Fig. 1a dashed green and black curves) agrees well with that derived from the optical absorption of freely suspended graphene [13].

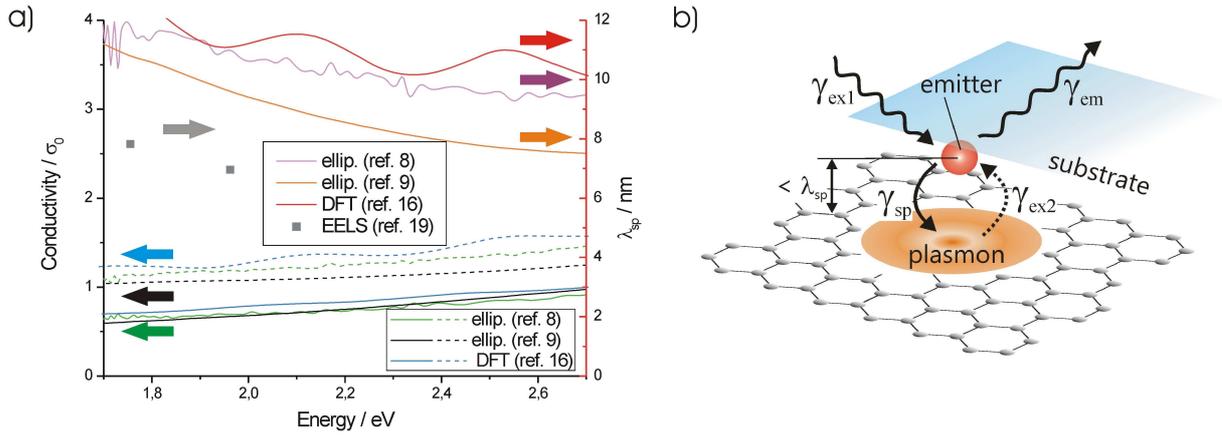

Figure 1. a) Left axis: Real and imaginary parts (dashed and solid curves, respectively) of the conductivity of graphene in units of the universal conductivity $\sigma_0 = e^2/4\hbar$ [11] calculated (equation 2) from the experimental (green and black curves) and theoretical (blue curves) refractive indices. Right axis: plasmon dispersion (equation 1) for graphene on a solid substrate (here $\varepsilon = 2.6$) calculated from the imaginary part of the conductivity (pink, orange and red curves). Grey squares are plasmon wavelengths for graphene on mica recalculated from the plasmon wavelength for graphene on silicon carbide (SiC) provided by EELS with respect to the dielectric constants of SiC (6.9) [14] and mica (2.6) [15]. b) Diagram of the experiment: graphenes were exfoliated onto a mica substrate covered with a submonolayer of R6G molecules. Direct excitation of graphene plasmons by the far-field ($\gamma_{ex1}$) is not possible due to the large wavevector mismatch. However, fluorescent molecules in close proximity to graphene can efficiently excite graphene plasmons ($\gamma_{sp}$), since they provide the large wavevectors, existing in the near-field of the emitters, necessary for graphene plasmon excitation. Excited graphene plasmons can subsequently re-excite the emitters ($\gamma_{ex2}$).

Furthermore, the complex conductivity of graphene is in good agreement with predictions by density functional theory (DFT) (see Fig. 1a), which in contrast to TBM and RPA is not only dominated by electronic transitions in the vicinity of the *K* points, but also includes significant contributions from electronic transitions along the *ΓK* and *ΓM* directions.[12,16] Therefore DFT calculations support graphene plasmons in the visible spectral range for undoped graphene as well as doping induced graphene plasmon enhancement also at plasmon energies well above the Fermi level shift [17], in strong contrast to TBM based RPA [3].

The graphene plasmon dispersion is obtained by substitution of graphene's experimental conductivities into equation 1 yielding graphene plasmon wavelengths on the order of 10 nanometers in the visible spectral range (Fig. 1a orange and pink curves), about 50 times below the wavelength of free space photons. Furthermore, we expect the plasmon dispersion in the visible to converge to the graphene plasmon resonance in the ultraviolet [18].

The graphene plasmon dispersion curves derived from graphene's measured optical properties is in good agreement with investigations of graphene by electron energy loss spectroscopy (EELS) [19,20] (see Fig. 1a). Additionally the results of EELS including higher order graphene plasmon modes in the ultraviolet [21] are in good agreement with DFT simulations [17].

The considerations above also imply graphene plasmons being optically excitable in the visible spectral range. However, efficient excitation of SPs requires matching of energy and momentum of the photons to the SPs [2], challenging due to the high graphene plasmon wavevectors. Yet graphene plasmons were predicted to be excitable by the large momenta existing in the near-field of a small emitter [2] located at a distance to graphene smaller than the graphene plasmon wavelength [3] of around 10 nm at optical frequencies.

In the following we present results on coupling of molecular emitters to graphene plasmons. We exfoliate graphenes onto transparent mica substrates covered with a submonolayer of rhodamine 6G molecules (R6G) to bring graphenes and the molecules in a subnanometer distance to each other (Fig. 1b). Steady state and time resolved fluorescence measurements under ambient conditions were used to quantify the coupling efficiency of the emitters and graphene plasmons, i.e. to quantify the fluorescence quenching efficiency of graphene (reduction of the emission rate $\gamma_{em}$).

Fig. 2a shows a fluorescence image of graphenes exfoliated onto a submonolayer of R6G on a mica substrate (see methods in the Supplementary Information and see ref. [22] and Supplementary Figure S1 for identification of graphenes) [23]. Graphene covered regions (I) appear darker in comparison to the bright fluorescence emitted from the uncovered R6G (X), which is attributed to quenching of the R6G fluorescence by the graphenes. Single layer graphene exhibits an apparent quenching factor of $q \equiv I_m/I_g = 5.2\pm0.7$, with $I_m$ and $I_g$ being the fluorescence intensities of uncovered and graphene covered areas respectively, i.e. the mean of the intensity histograms in Fig. 2b (quoted error is the standard deviation of the mean from different samples). We observed only a weak increase of the quenching factor with the number of graphene layers: e.g., 6 ± 1 layers of graphene quench the fluorescence by a factor of 8.2 ± 1.2 (not shown). However, the apparent fluorescence quenching of R6G by graphenes is rather a lower limit of the quenching factor: first of all scanning force microscopy investigation of the graphene covered R6G submonolayers have shown that R6G is confined flatly by graphene (transition dipole moment parallel to graphene) while uncovered cationic dye molecules are known to adsorb rather disordered on mica, which is also reasonable for (cationic) R6G. This and an additional possible red shift of the R6G absorption in the graphene region results in a higher excitation and collection efficiency of R6G in the graphene region. Correcting the apparent quenching factor for equal excitation and emission detection conditions of the R6G fluorescence in the mica and graphene regions we obtain an effective quenching factor $q_{eff}$ between 16 and 181 (see Supplementary Equations S1 – S7) [23].

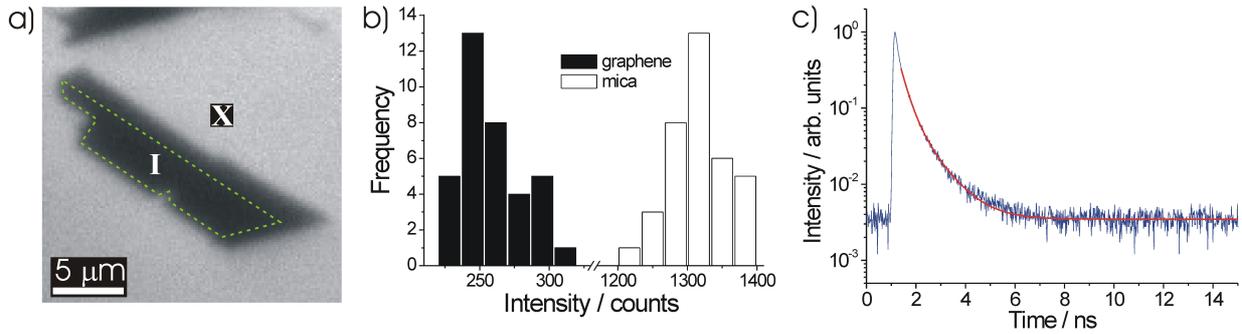

Figure 2. a) Fluorescence microscopy image of graphenes on R6G covered mica substrate. Single layer graphene is outlined with the green dashed line and marked with (I), regions not covered by graphene are marked with (X). b) Fluorescence intensity histograms from single layer graphene covered (I, black bars) and uncovered (X, white bars) areas, respectively. Fluorescence of molecules under single layer graphene is significantly quenched, but still detectable. c) Time resolved fluorescence decay of R6G on mica excited at 532 nm (blue curve). A double exponential decay fits the data (red curve). The resulting fluorescence life times are 0.28 ± 0.02 ns and 0.89 ± 0.05 ns with the amplitude of the faster decay being more than one order of magnitude stronger.

In the following we will compare the results on measured quenching through graphene to that expected from theory. The effective quenching factor equals the ratio of the R6G fluorescence emission rates (intensity) in the mica and graphene regions $\gamma_{em}^s$ and $\gamma_{em}^g$, respectively:

$$q_{eff} \equiv \frac{\gamma_{em}^s}{\gamma_{em}^g}. \tag{3}$$

The fluorescence rate in the mica region can be written as (see also ref. [24]):

$$\gamma_{em}^s = \gamma_{ex}^s / (\gamma^s / \gamma_r), \tag{4}$$

with $\gamma_{ex}^s$ being the excitation rate, $\gamma_r$ the radiative rate and $\gamma^s$ the total decay rate in the mica region. Accordingly the fluorescence rate in the graphene region can be written as:

$$\gamma_{em}^g = \gamma_{ex}^g / (\gamma^g / \gamma_r). \tag{5}$$

Substitution of equation 4 and 5 into equation 3 and rearranging the same gives

$$(\gamma_{ex}^g / \gamma_{ex}^s) / (\gamma^g / \gamma_r) = \gamma_r / (q_{eff} \, \gamma^s) \equiv \gamma_{em}^{norm}, \tag{6}$$

which we refer in the following to as the normalized fluorescence rate $\gamma_{em}^{norm}$. $(\gamma_{ex}^g / \gamma_{ex}^s)$ deviates from unity only if local excitation enhancement is present (e.g. in the case of plasmon active metal surfaces) [24]. The normalized fluorescence rate relates the radiative rate to the total decay rate induced by graphene and is thus a proper quantity to compare quenching measured in different experimental configurations (e.g. varying substrate induced decay). Therefore we will convert the measured effective quenching factor (see above) into the normalized fluorescence rate using equation 6. For this $\gamma_r$ and $\gamma^s$ are required.

In order to estimate the total decay rate $\gamma^s = 1/\tau^s$ of R6G on mica, with $\tau^s$ being the measured dominating fluorescence lifetime, time resolved fluorescence decay measurements (Fig. 2c) were performed (see methods in the Supplementary Information) [23]. A double exponential decay was found to fit the data, similarly as for rhodamines on other substrates

[25]. The resulting fluorescence life times are $0.28 \pm 0.02$ ns and $0.89 \pm 0.05$ ns with the amplitude of the faster decay being more than one order of magnitude stronger.

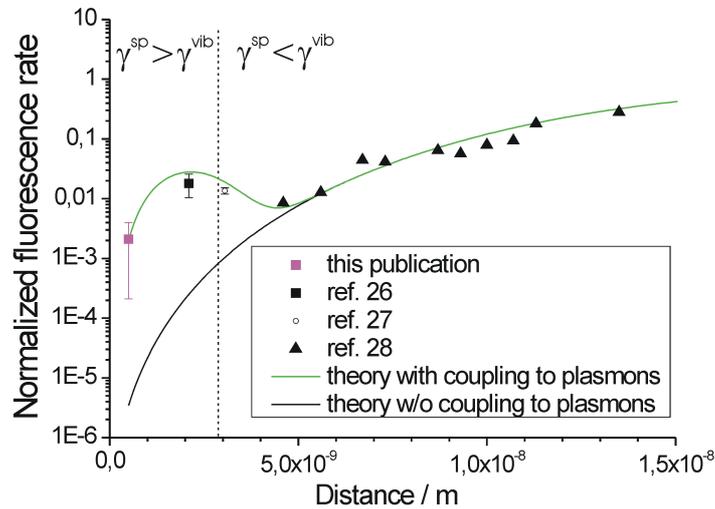

Figure 3. Distance dependent normalized fluorescence rate with (green curve) and without (black curve) excitation enhancement. A striking deviation of the experimental data to the theory without plasmon excitation at distances smaller than the graphene plasmon wavelength is evident. The pink square is the normalized fluorescence rate measured in our experiment. Black square, circle and triangles are normalized fluorescence rates obtained from the measured quenching factors and rates adopted from ref. [26], ref [27] and ref. [28], respectively. Note that the curves are calculated for the experimental parameters given in this publication. The experimental parameters (emission wavelength, substrate refractive index) in ref. [26], ref. [27] and ref [28] differ somewhat, the resulting deviations in the calculations are, however, insignificant for the comparison to theory with and without plasmon. Ratios of excited state relaxation rate to radiative rate, necessary for calculation of the excitation enhancement (equation 9), are also comparable to R6G (see ref. [27,29-33]).

The dominating lifetime of R6G on mica is about one order of magnitude shorter compared with lifetimes of rhodamines in solutions [31] which can be attributed to an increase of the non-radiative decay rate resulting from interaction with the substrate [25]. Substrate interaction is also indicated by fluorescence spectra of R6G on mica (see Supplementary Figure S3) [23].

Assuming the radiative decay rate $\gamma_r$ for R6G in the mica and graphene regions under ambient conditions being equal to the radiative decay rate of R6G in water of $2.2 \times 10^8$ s$^{-1}$ (ref. [31]) and with $1/\gamma^s = 0.28 \pm 0.02$ ns and $q_{eff}$ between 16 and 181 (see above) this yields according to equation 6 a measured normalized fluorescence rate between $3.2 \times 10^{-4}$ and $4.1 \times 10^{-3}$. This experimental result will now be compared with theoretical predictions.

The normalized total decay rate $\gamma^g/\gamma_r$ in the graphene region (including excitation of plasmons and electron hole pairs) for an emission dipole parallel to the graphene surface (like in our experiment) [23] can be calculated from graphenes optical conductivity $\sigma$ at the emission frequency $\omega$ of the emitter [3] (emission maximum at about 535 nm, see Supplementary Figure S3) [23]:

$$\gamma^g/\gamma_r = 1 + \frac{3c^3}{2\omega^3} \int_0^\infty k^2 dk \, \text{Im}\left\{\frac{-1}{\varepsilon + 1 + 4\pi i k \sigma/\omega}\right\} e^{-2kz}, \qquad (7)$$

where $z$ is the distance between emitter and graphene, $\varepsilon$ is the dielectric constant of the substrate and $c$ the speed of light.

With an optical conductivity $\sigma = 1.2+0.8i$ (average value at 535 nm, see Fig. 1a), the dielectric constant of mica $\varepsilon_{mica} = 2.6$ (ref. [15]) at 535 nm and a distance between R6G and graphene of $z = 0.5$ nm (center-to-center distance of graphene and R6G, estimated from the SFM images of R6G confined by graphene and the thickness of graphene, see Supplementary Figure S2) [23], a normalized total decay rate $\gamma^g/\gamma_r$ of $2.9 \times 10^5$ is calculated. Assuming no local excitation enhancement this value is identical to the inverse of the normalized fluorescence rate $\gamma_{em}^{norm}$ and thus yields a normalized fluorescence rate of $3.5 \times 10^{-6}$. Also usage of the equations for the non-radiative decay rate based on the theoretically calculated dielectric properties of graphene (TBM [34] and TBM based RPA [35], no plasmon activity in the visible) results in normalized fluorescence rates close to the ones calculated above from the measured dielectric properties with the equation published in ref. [3]. Note, that the calculated normalized fluorescence rates are between two and three orders of magnitude smaller than the measured one. Possible charge transfer quenching, which is not taken into account by the equations above, results only in a further decrease of the calculated normalized fluorescence rate and can thus only increase the discrepancy between calculated and measured normalized fluorescence rates.

However, so far we did not take into account excitation enhancement provided by graphene plasmons. Similar to plasmons at metal surfaces the graphene plasmon contribution to the total decay rate of the emitter is estimated from the pole contribution of the local reflected field.[36,37] In particular the graphene plasmon contribution to the normalized decay rate in the graphene region $\gamma^{sp}/\gamma_r$ can be calculated from the graphene plasmon wavelength $\lambda_{sp}$, which is related to the optical conductivity of graphene (see Fig. 1a) at the emission frequency $\omega$ of the emitter [3]:

$$\gamma^{sp}/\gamma_r \approx \frac{12\pi^4 c^3}{(\varepsilon+1)\omega^3} \frac{e^{-4\pi z/\lambda_{sp}}}{\lambda_{sp}^3}, \tag{8}$$

where $z$ is the distance between emitter and graphene, $\varepsilon$ is the dielectric constant of the substrate and $c$ the speed of light. Substitution of an optical conductivity $\sigma$ = 1.2+0.8i at 535 nm and the dielectric constant of mica (see above) into equation 1 gives a plasmon wavelength $\lambda_{sp}$ of 9 nm. The imaginary part of the plasmon wavevector reflects the in-plane propagation distance [3] $l_{sp}$ (here equal to 6 nm) and indicates that the plasmon is damped in the visible spectral range due to generation of electron-hole pairs, but is still capable of field enhancement as we will show in the following.

By substituting the graphene plasmon wavelength and the same values as above into equation 8 the plasmon contribution to the normalized decay rate of the emitter $\gamma^{sp}/\gamma_r$ is calculated to 1.4x10$^5$, which is on the same order as the normalized total decay rate (see above). Thus, the plasmon-related decay is a major contribution to the total decay. This is changing at emitter-graphene distances on the order of the graphene plasmon wavelength, at which the emitter can not support the large wavevectors needed to excite plasmons anymore and thus also direct excitation of electron-hole pairs becomes significant (see Fig. 3 and discussion below).[3]

Excitation enhancement is caused by excited graphene plasmons that subsequently re-excite, not necessarily coherent, the emitters (also called self-reaction or secondary field) [2] (see cartoon in Fig. 1b). As a result in the graphene regions an increase of the normalized fluorescence rate is observed. The excitation enhancement ($\gamma_{ex}^g/\gamma_{ex}^s$) scales with the square of the enhancement of the electric field at the position of the emitter [2] (as excitation of graphene plasmons by free space photons is negligible). Since the field enhancement results from the confinement of the photon field with a wavelength $\lambda_0$ to the field of the graphene plasmons with a wavelength $\lambda_{sp}$, we can estimate the magnitude of the excitation enhancement at the origin of the plasmon from the reduction of the mode volume given by $(\lambda_0/\lambda_{sp})^2$ [3] yielding a correspondent increase of the energy density [38] (equal to the square of the field enhancement). The electric field normal to graphene exponentially decays as exp(-$z$ Re($k_{sp}$)) and in the plane of graphene as exp(-$x$ Im($k_{sp}$)) due to damping of the plasmon away from the point of its excitation by generation of electron-hole pairs in graphene.[3] Assuming that the extension of the plasmon is large compared to that of the absorption dipole of the emitter, which is justified here, the excitation enhancement by plasmon-mediated re-excitation of R6G in the graphene covered areas relative to the uncovered mica areas can be written as:

$$\gamma_{ex}^{g}/\gamma_{ex}^{s} = 1 + g_{sp \to em} g_{em \to sp} \frac{\lambda_0^2}{\lambda_{sp}^2} e^{-4\pi z/\lambda_{sp}}. \qquad (9)$$

Similar as in ref. [36] we introduce as additional correction first the coupling efficiency of the emitters to graphene surface plasmons $g_{em \to sp} = \gamma^{sp}/(\gamma^{sp} + \gamma^{g})$ to include reduction of the coupling efficiency to graphene plasmons through direct excitation of electron-hole pairs in graphene. Second, with the coupling efficiency of graphene plasmons to the emitters $g_{sp \to em} = \gamma^{sp}/(\gamma^{sp} + \gamma^{vib})$, we include that efficient re-excitation is only possible if the decay rate into graphene plasmon $\gamma^{sp}$ (here $3.1 \times 10^{13} s^{-1}$, see above) is larger than the vibrational relaxation rate $\gamma^{vib}$ of the emitter [2] ($\approx 10^{12}$ s$^{-1}$ for R6G) [33]. The latter is equivalent with a negligible spectral red shift between the energy of the graphene plasmon and the absorption maximum of the emitter, i.e. a negligible spectral mismatch (indicated by the dashed vertical line in Fig. 3). Furthermore a damping induced spectral broadening of the graphene plasmon which can also result in a reduced spectral overlap of graphene plasmon and emitter is considered in the following.

The plasmon linewidth $\Gamma$ is related to the plasmon group velocity $v_{sp}$ (in graphene equal to the Fermi velocity $10^6$ m/s) [39] and its propagation distance $l_{sp}$ (6 nm, see above) by $\Gamma = (\hbar v_{sp})/l_{sp}$ (ref. [37]) and can thus be estimated to 0.1 eV. Note that the plasmon group velocity can not be estimated from the slope of the plasmon dispersion (see Fig. 1a) in the case of anomalous dispersion (which is given here) due to wave profile deformation. Instead it coincides with to the plasmon group velocity in the normal dispersion region as used above.[40] A further estimation can be conducted by considering broadening of the graphene plasmon linewidth through dephasing [41] (damping) of the collective oscillation of electrons. Taking the lower limit of the nonequilibrium (quasi free) carrier relaxation time in graphene (0.01 ps) [42] as an estimate for the graphene plasmon dephasing time $\tau_{sp}$, one can estimate an upper limit of the graphene plasmon linewidth $\hbar/\tau_{sp}$ of even less than 0.1 eV. Since these value are on the same order as the full width half maximum of the R6G absorption [43] we can neglect reduced spectral overlap through spectral broadening of the graphene plasmon in our calculations. Note that structural defects in epitaxial graphene can lead to larger graphene plasmon linewidths,[39] these defects are, however, unreasonable for the exfoliated graphenes used here.

Using the values for $z$, $\lambda_{sp}$ and $\lambda_0$ given above yields a relative excitation enhancement $\gamma_{ex}^g/\gamma_{ex}^s$ in the graphene covered sample regions on the order of 600, nearly 3 orders of magnitude. Substitution into equation 6 yields a normalized fluorescence rate on the order of $2 \times 10^{-3}$, which is in the same range as the measured normalized fluorescence rate in contrast to the underestimation by orders of magnitude when neglecting possible re-excitation due to plasmons.

Figure 3 contains the key result of our paper and displays the normalized fluorescence rate for emitters as a function of the distance to graphene (green curve with and black curve without plasmon induced excitation enhancement). The graph implies that the difference between including and excluding excitation enhancement is only pronounced at distances smaller than the graphene plasmon wavelength at which efficient graphene plasmon excitation and strong field enhancement is present, similarly as demonstrated for plasmons at metal surfaces [24,44]. The calculated normalized fluorescence rate is in very good agreement with fluorescence rates of emitters in proximity to graphene measured at larger distances in our lab [26] (fluorescent polymer film with a typical quantum yield of 0.03±0.01 [32]) and by others [27,28]. Note, that not a single fitting parameter is used in our model, on the contrary the model fits naturally to the data.

Quenching by multilayer graphene can be calculated by treating every layer of graphene in a multilayer graphene sample as an individual decay channel (superposition) with the individual layers separated by 0.34 nm as suggested in ref. [27]. Including in this case also the excitation enhancement induced by each layer the normalized emission rate for R6G confined by a 6-layer graphene can be obtained by summing up the total decay rate normalized by the excitation enhancement of each layer which gives a normalized emission rate of $1.2 \times 10^{-3}$. This value is 1.7 times smaller than the normalized emission rate estimated for single layer graphene, which agrees well with the ratio of the experimental quenching factors for single and 6-layer graphene samples, respectively (see above). Thus our data also indicate emitter-plasmon coupling for multilayer graphene. Similarly as for single layer graphene one would expect a decrease of the normalized fluorescence rate by orders of magnitude when coupling to plasmons is neglected (see above). In addition this good agreement further supports the validity of our model visualized in Fig. 3 for graphene-emitter distances up to 2.2 nm (distance between R6G and the last layer of a 6-layer graphene). Note, that plasmon activity of few and multilayer graphenes is also indicated by EELS.[20,21]

In conclusion, we demonstrated graphene plasmon activity in the visible spectral range. Its existence is implied by the optical conductivity of graphene and we also gave direct evidence by quantifying the quenching efficiency of graphene. These results agree well with EELS experiments and DFT predictions (see Fig. 1a). Plasmonic activity in the visible introduces graphene not only as an exciting material for plasmonic devices for technological application at optical frequencies, but also its understanding is mandatory for the interpretation of graphene based optical experiments (e.g. graphene induced Raman enhancement) [7]. The extraordinarily strong confinement of graphene plasmons, implied by their short wavelength and short propagation distance and thus the outstanding emitter-plasmon coupling strength, despite strong intrinsic plasmon damping, suggests graphene as a candidate for highly integrated nano-optoelectronic devices. Future experiments should address graphene plasmon tuning in the visible spectral range.

Acknowledgements:

Financial support by SFB 951 "Hybrid Inorganic/Organic Systems for Opto-Electronics" and by the Integrative Research Center for the Sciences IRIS Adlershof are gratefully acknowledged.

# Supplementary Information

# Evidence for graphene plasmons in the visible spectral range probed by molecules

*Philipp Lange, Günter Kewes, Nikolai Severin, Oliver Benson, Jürgen P. Rabe*

Humboldt-Universität zu Berlin, Department of Physics, 12489 Berlin, Germany

**Contents**

I. Methods

**Sample preparation.** Submonolayers of rhodamine 6G (R6G) were prepared by spincoating from aqueous solutions (purified, 0.1 µm filtered water, Sigma-Aldrich, concentrations 0.03 – 0.1 mM) onto freshly cleaved muscovite mica (Ratan mica exports, V1 (optical quality)) under ambient conditions (22-23°C, 40 - 48% relative humidity (RH)). The rear side of the substrate was cleaved off after spincoating to remove its possible contaminations. Submonolayer thickness of R6G was confirmed by absorption spectroscopy and scanning force microscopy (see Supplementary Figures S2 and S6).

Graphenes were deposited onto the R6G covered mica substrates by mechanical exfoliation of highly oriented pyrolytic graphite (Momentive, HOPG, grade ZYB). The HOPG bulk crystal was cleaved with scotch tape prior to the exfoliation process. Then, the freshly cleaved surface of the HOPG bulk crystal was brought gently into contact with the R6G covered mica substrates.

**Fluorescence measurements.** Fluorescence images were acquired with an Axiovert 100 TV equipped with a HBO 50 microscope lamp with UV and infrared blocking filters (Zeiss) and a Plan 40x/0.55 object lens (Nikon) and a cooled CCD camera (SC4022, EHD). The samples were excited and the subsequent emission was detected through the mica substrates (445 nm - 565 nm band pass excitation filter (mainly transmitting the 546 nm peak of the HBO lamp), 560 nm dichroic filter, 585 nm long pass emission filter, Laser Components). The intensity of images recorded from mica substrates, not covered with R6G and freshly cleaved from both sides, in the same setup used to record fluorescence images was regarded as background and subtracted from the fluorescence intensity of the R6G covered mica substrates.

Time resolved fluorescence decay measurements of R6G on mica substrate were performed with a confocal microscope based on an inverted fluorescence microscope (Axiovert 200, Zeiss). The dyes were excited at 532 nm (LDH-P-FA-530, PicoQuant), and the fluorescence was detected through a longpass filter with an avalanche photodiode (SPCM-AQR-14, Perkin Elmer); the time intervals between excitation and detection of fluorescence were measured by time-correlated single photon counting (Picoharp 300, PicoQuant).

## II. Identification of graphenes

Graphenes were identified optically by reflection microscopy,[1] using an Axiovert 100 TV equipped with a HBO 50 microscope lamp with UV and infrared blocking filters (Zeiss) and a Plan 40x/0.55 object lens (Nikon). The samples were illuminated through the mica substrate (lamp spectrally narrowed with a 565 nm – 595 nm band pass filter), the reflected light was detected through crossed polarizers. Images were acquired with a cooled CCD camera (SC4022, EHD).

Fig. S1 shows a reflection microscopy image of graphenes on a submonolayer of R6G on a mica substrate. Graphenes appear dark on the mica substrate. The region marked with (I) exhibits a contrast of $10.5\% \pm 0.3\%$ on the background (X) and can thus be identified as single layer graphene.[1]

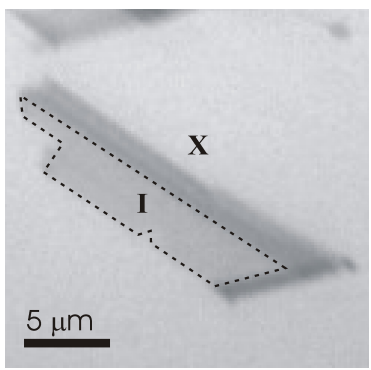

Figure S1. Reflection microscopy image of graphenes on R6G covered mica substrate. Single layer graphene is outlined with the black dashed line and marked with (I), regions not covered by graphenes with (X).

### III. Scanning force microscopy of R6G confined by graphenes

As graphene is known to follow the substrate topography even on a molecular scale [2] scanning force microscopy (SFM) was used to confirm the assumed submonolayer thickness of R6G confined by graphene and to investigate the orientation of the confined molecules. This information is necessary to quantify effective fluorescence quenching by graphene (see below and main part of the manuscript).

The SFM images were recorded by operating in tapping mode under controlled humidity (MultiMode SFM, Veeco Metrology; Nanoscope IV SFM controller, Digital Instruments; AC160TS SFM cantilever, Olympus). The SFM head was placed in a bell jar chamber; the humidity inside the chamber was measured with a humidity sensor (testo 625, Testo AG) and controlled by purging with dry nitrogen or dry nitrogen bubbled through purified water.

Fig. S2a shows a SFM image of double layer graphene exfoliated onto R6G covered mica substrate (see sample preparation) and imaged at 30% relative humidity (RH). The topography exhibits plateaus and small elevations with a height of 0.34 ± 0.05 nm. The full width half maximum of the small elevations is about 10 nm. It is noticeable that small elevations are rarely located between laterally larger plateaus. The surface density of the small elevations in the regions not exhibiting larger plateaus is on the order of $2 \times 10^3$ per µm². Since graphene exfoliated onto mica substrate is known to be atomically flat under ambient condition [3] the elevations and plateaus on the SFM image (Fig. S2a) are attributed to R6G molecules confined between mica and graphene. The fact that the small elevations are rarely located in between the laterally larger plateaus indicates that the plateaus consist of laterally closely packed elevations.

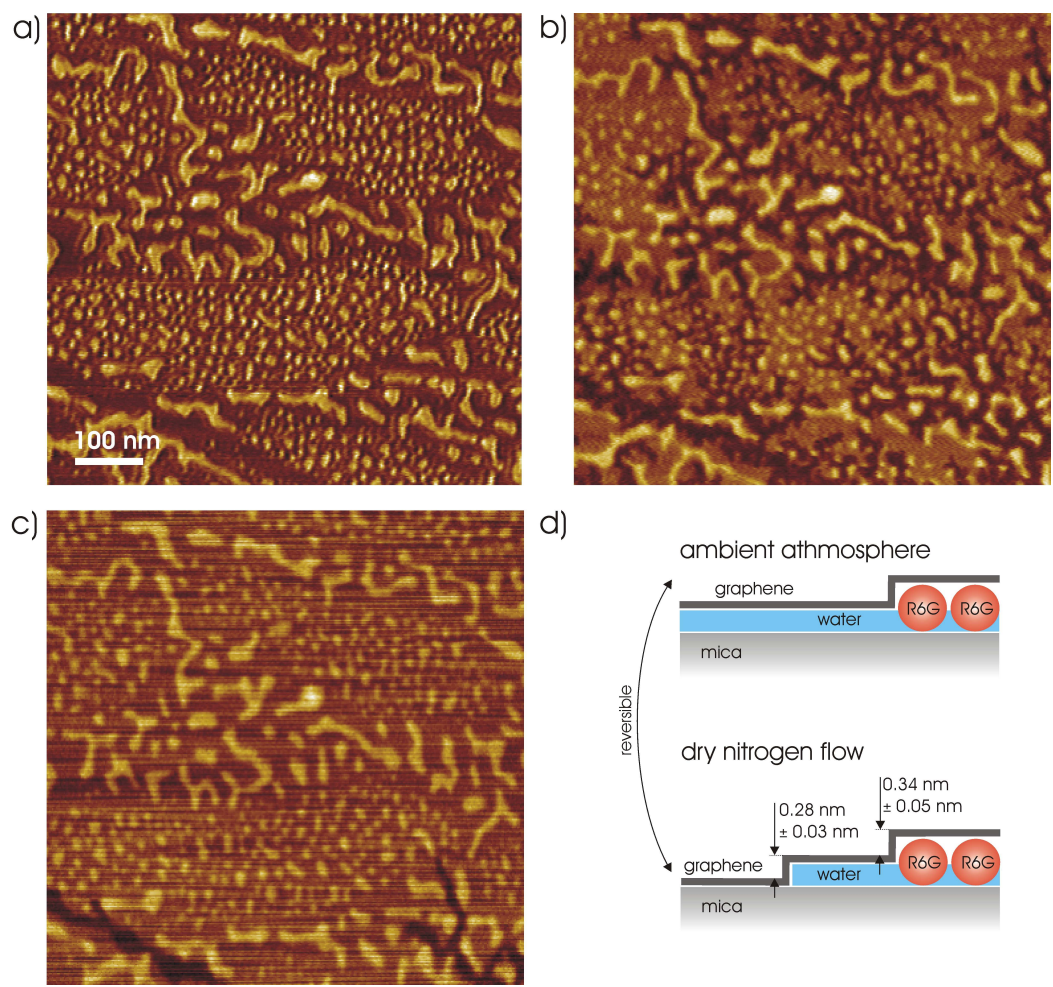

Figure S2. SFM images of double layer graphene exfoliated onto a R6G covered mica substrate (prepared from 0.1 mM) imaged at a) 30% RH, b) 1.5% RH after continuous drying for half an hour and c) 55% RH after continuous rising the humidity for the quarter of an hour. d) Schematics of the proposed sample structure before and after dewetting of the water layer, indicated heights are the heights measured by SFM (a and b).

As a molecularly thin water layer is known to be confined between mica and graphene under ambient conditions, the regions between the elevations and plateaus should therefore contain water. This water layer should be dewettable upon lowering the humidity [3], which we will check in the following.

Fig. S2b displays the same regions as in a) after continuous purging with dry nitrogen for half an hour and finally imaged at 1.5% RH. Some areas in the regions between the elevations and plateaus now exhibit depressions with a depth of 0.28 ± 0.03 nm. Rising the humidity results in disappearance of the depressions (Fig. S2c, imaged at 55% RH after continuous rising the humidity for the quarter of an hour). Thus the depression growing between the elevations and plateaus upon lowering the humidity are attributed to result from dewetting of the molecularly thin water layer confined between mica and graphene additionally to the R6G molecules. The measured depth of the depressions (0.28 ± 0.03 nm) fits well to the depth of the depressions measured on graphene samples not containing R6G molecules additionally to water.[3]

By adding the height of the water layer to the height of the elevations (molecules) above the water layer we estimate the total apparent height of the R6G molecules to be 0.62 ± 0.08 nm (see Fig. S2d). As the thickness of a R6G molecule perpendicular to its plane is calculated to range between 0.5 nm and 0.85 nm (see ref. [4]) we can conclude that the molecules are approximately flatly confined between mica and graphene. Consequently we assume that the dipole moment of R6G is oriented approximately parallel to the graphene layer.

Assuming that the measured full width half maximum of the elevations corresponds approximately to the lateral extension of the elevations, that the R6G molecules are densely packed within the elevations and that the cross sectional area of a single flatly oriented R6G is 1.56 nm$^2$ (see ref. [4]) one can estimate a R6G surface density on the order of $10^5$ molecules per μm$^2$. From the SFM measurements one can not exclude that some molecules are also located on top of graphene. However, since the estimation of the R6G surface density in the graphene and mica regions by contrast (not shown) and absorption spectroscopy (see below), respectively, yield R6G surface densities on the same order as by SFM (see above), we conclude that the R6G surface densities in the mica and graphene regions are on the same order.

### IV. Fluorescence spectra of R6G on mica

Fluorescence spectra of R6G on mica were recorded to verify the fluorescence maximum of R6G on mica under ambient conditions since this is a parameter needed for the calculation of the quenching factor. Additionally the spectra provide evidence that the short fluorescence lifetime of R6G on mica results from interaction of R6G with the mica substrate.

The fluorescence was excited from 450 nm – 490 nm, detected through a 515 nm long pass filter and recorded with a CCD spectrometer (SP-150 spectrograph, Acton Research Corporation, equipped with a 300 grooves/mm grating and a LN/CCD-1340/100-EB/1 detector, Roper Scientific).

Fig. S3 shows fluorescence spectra of R6G on mica substrate at different relative humidities and of R6G dissolved in water. The fluorescence spectrum of R6G on mica is even at high relative humidities (maximum at 544 nm at 81% RH) blue shifted (9 nm) with respect to the solution spectrum (maximum at 553 nm) (see also ref. [5]). Interestingly the blue shift of the spectrum is increased by reducing the relative humidity of the surrounding environment (e.g. to about 535 nm at 48% RH, the humidity range at which the fluorescence images in the main part of the manuscript were recorded).

The observed blue shift of the fluorescence spectrum of R6G on mica could result from the lack of a solvent induced bathochromic shift (also observed for R6G in the gas phase)[6] and interaction of R6G with the mica substrate. Possible interaction mechanisms with the substrate are formation of a contact ion pair [7], ion exchange with the mica surface [8] and charge transfer with the substrate [9]. Since the fluorescence of R6G on mica at high humidities is still blue shifted with respect to the fluorescence of R6G dissolved in water, the blue shift is attributed to a superposition of interaction with the mica substrate and the lack of a bathochromic shift.

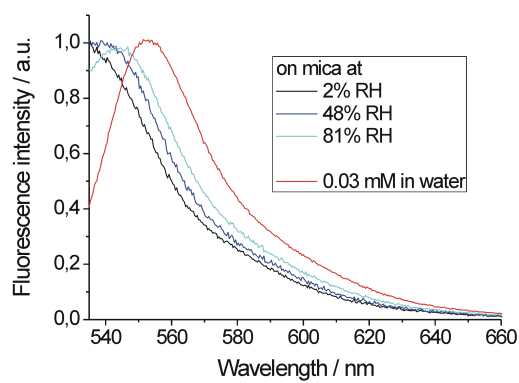

Figure S3. Fluorescence spectra of R6G in water (0.03 mM, red curve) and of R6G on mica surface spincoated from 0.03 mM solution at varying humidities (2% RH, black curve; 48% RH, blue curve; 81% RH, light blue curve).

## V. Excitation efficiency of R6G in the mica and graphene regions

In this section we will estimate the impact of the excitation efficiency of R6G in the mica and graphene regions on the apparent quenching factor. The excitation efficiency of R6G in the mica and graphene regions might vary due to different orientations of R6G and spectral shifts of the R6G absorption in the mica and graphene regions which both can affect the absorbed power. First we will discuss the orientations of R6G in the mica and graphene regions.

Using SFM we have shown that R6G is confined flatly between mica and graphene (see above). However, cationic dye molecules (crystal violet and malachite green) are known to adsorb on mica with a maximum angle between the plane of the molecule and the mica substrate of about 60° or rather unordered (average tilt angle of 45°).[8] These orientation are also reasonable for (cationic) R6G on mica, since for this the estimation of the R6G surface density from absorption spectroscopy measurements (see below) of uncovered R6G on mica is in good agreement with the estimation from SFM of R6G confined between mica and graphene (see above). Consequently the different orientation of R6G in the regions covered and not covered by graphene might influence the apparent quenching factor.

In the following we will estimate the dependency of the excitation efficiency of R6G on the illumination aperture and the orientation of the R6G absorption dipole.

The samples were illuminated through the object lens. Since the microscope used for fluorescence imaging is not equipped with an aperture stop, the illumination aperture and accordingly the maximum illumination angle $\alpha$ is equal to the aperture of the object lens ($NA = 0.55$, $\alpha \cong 33°$). This can be described as an effective illumination angle $\alpha' \cong 24°$, which we will show in the following.

We assume the intermediate image of the light source to be a uniform bright round area with radius $r$ and the circles around its centre being linearly distributed onto the illumination angles (only a point source will result into NA = 0). Furthermore, using that the centre $r'$ of an area of a segment of a circle with radius $r$ is equal to $r' = r/\sqrt{2}$ and that $\alpha/r$ is the optical invariant [10], the effective illumination angle can be estimated to $\alpha' = \alpha/\sqrt{2} \cong 24°$ (Fig. S4a).

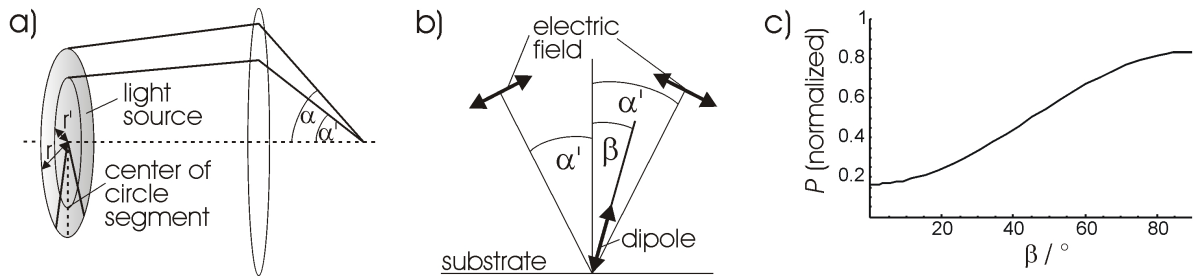

Figure S4. a) Sketch for the estimation of the effective illumination angle $\alpha'$ from the maximum illumination angle $\alpha$ (corresponding to the illumination aperture) and the radius of the intermediate image of the light source $r$ using that $\alpha/r$ is the optical invariant of the system. b) Sketch for the calculation of the illumination aperture and dipole orientation dependent absorption, with $\beta$ being the angle between the absorption dipole and the optical axis and $\alpha'$ the effective illumination angle. c) Dependency of the absorbed power $P$ on the dipole orientation $\beta$ for an illumination aperture $NA = 0.55$.

To estimate the dependency of the excitation efficiency of R6G we now use that the power absorbed by a dipole is proportional to the square of the electric field projected onto the dipole [11] (Fig. S4b), which can be written as:

$$P \propto |\sin(\alpha'-\beta)|^2 + |\cos(90-\alpha'-\beta)|^2, \tag{S1}$$

with $\alpha'$ being the effective illumination angle and $\beta$ being the angle between the axis of the dipole and the optical axis. A dipole parallel to the optical axis will absorb 0.17, 45° to the optical axis 0.5, 30° (60° to the substrate) to the optical axis 0.33 and 90° to the optical axis 0.83 of the maximum possible absorbed power (Fig. S4c). From this one can estimate that the maximum orientation dependent variation of the absorbed power is smaller than a factor of 0.83/0.17 ($\cong 4.9$) and in particular that corresponding to the orientation of R6G under graphene a dipole parallel to the substrate will absorb 1.7 times more than a dipole tilted 45° and 2.5 times more than a dipole tilted 30° to the optical axis (equal to a tilt of 60° to the substrate), corresponding to the orientations of R6G on mica.

Additionally the absorption maximum of R6G in the graphene regions might be red shifted [12] to the absorption of R6G in the mica region (see below) and could consequently be closer to the used emission maximum of the microscope lamp. This implies an up to three times higher excitation efficiency of R6G in the graphene region.

The apparent quenching factor needs to be multiplied by the discussed factors to provide equal excitation conditions in the mica and graphene regions, i.e. effective quenching is here higher than apparent quenching.

### VI. Collection efficiency of the R6G fluorescence in the mica and graphene regions

Besides the excitation efficiency (see above) also the collection efficiency of the R6G fluorescence in the mica and graphene regions depends on the orientation of R6G and can thus impact the apparent quenching factor.

In the following an equation for the estimation of the collection efficiency of the object lens is derived, related to the numerical aperture of the object lens, the orientation of the R6G emission dipole and the transmittance of the mica substrate.

The power of a dipole in a homogeneous environment $P(\vartheta,\varphi)$ radiated into an infinitesimal small unit solid angle $d\Omega = sin(\vartheta)d\vartheta d\varphi$, expressed in spherical coordinates with the dipole pointing along the z-axis and normalized by the total radiated power $\bar{P}$ is given by [11]:

$$\frac{P(\vartheta,\varphi)}{\bar{P}} = \frac{3}{8\pi}\sin^2(\vartheta). \tag{S2}$$

The fraction $\eta_c$ of the total emitted power transmitted through a substrate with the transmittance $T$ and collected by an object lens with an aperture angle $\alpha$ (maximum collection angle) smaller than the critical angle of total internal reflection (which is given here) can be calculated by integration of the equation above with respect to the transmittance [13]:

$$\eta_C = \frac{3}{16\pi}\left(\int_{-\pi}^{\pi}\int_0^{\alpha+\beta}T(\vartheta-\beta)\sin^3(\vartheta)d\vartheta d\varphi + l\int_{-\pi}^{\pi}\int_0^{\alpha-\beta}T(\vartheta+\beta)\sin^3(\vartheta)d\vartheta d\varphi\right), \tag{S3}$$

with $\beta$ being the angle between the optical axis and the dipole axis and $l = 1$ for $\alpha > \beta$ and accordingly $l = -1$ for $\alpha < \beta$.

The transmittance of the substrate is calculated from the Fresnel transmission coefficients of the air-mica and mica-air interfaces. Defining the plane of incidence being parallel to the dipole axis, the transmittance of the air-mica interface is related to the transmission coefficient $t_1^p$ for parallel polarized waves. Since mica is birefringent and thus the p-polarized wave is split into a p- an s-polarized part during propagation through the substrate, the transmission coefficients $t_2^p$ and $t_2^s$ for p- and accordingly s-polarized light need to be consider for the transmittance of the mica-air interface. As mica substrates with a thickness of at least several micrometers were used, we assume the light to be approximately unpolarized upon reaching the mica-air interface (equally distribution of p- and s-polarized light, as summation over dipoles with arbitrary orientation to the mica crystal axes parallel to the cleaved mica surface). The transmittance is thus estimated to:

$$T = \left( t_1^p \frac{(t_2^p + t_2^s)}{2} \right)^2, \quad (S4)$$

with the transmission coefficients $t_1^p$, $t_2^p$ and $t_2^s$ equal to:

$$t_1^p = \frac{2 \, n_{air} \cos(\gamma_1)}{n_{mica} \cos(\gamma_1) + n_{air} \cos(\gamma_2)}, \quad (S5)$$

$$t_2^p = \frac{2 \, n_{mica} \cos(\gamma_2)}{n_{air} \cos(\gamma_2) + n_{mica} \cos(\gamma_1)} \quad (S6) \text{ and } t_2^s = \frac{2 \, n_{mica} \cos(\gamma_2)}{n_{mica} \cos(\gamma_2) + n_{air} \cos(\gamma_1)}. \quad (S7)$$

The refraction index of air and mica is given by $n_{air} = 1$ and $n_{mica} = 1.596$, respectively and the angle of incidence $\gamma_1$ ranging from $\{0 \ldots \alpha\}$ and transmission $\gamma_2$ are related through Snell's law of refraction by $\gamma_2 = \arcsin(n_{air}/n_{mica} \sin(\gamma_1))$.

The numerical aperture of the object lens ($NA = 0.55$) yields a maximum collection angle $\alpha \cong 33°$. If the dipole axis is parallel to the substrate (corresponding to R6G under graphene,

$\beta = 90°$) a collection efficiency of 0.33 is obtained. An unordered orientation of the dipoles ($\beta = 45°$) or an angle between the substrate and the dipole axis of 60° ($\beta = 30°$) (corresponding to R6G on mica) yields a collection efficiency of 0.16 and 0.08, respectively. This means that the light emission from molecules with emission dipole parallel to the substrate surface is collected 2.1 times more efficient than from molecules with the emission dipole oriented 45° to the substrate surface and 4.1 times more efficient than from molecules with the emission dipole orientated 60° to the substrate surface.

The apparent quenching factor needs to be multiplied by the discussed factors to provide equal conditions for the collection of the R6G fluorescence in the mica and graphene regions. To obtain the effective quenching factor both the correction factors for equal conditions for the fluorescence collection and excitation (see above) must be included.

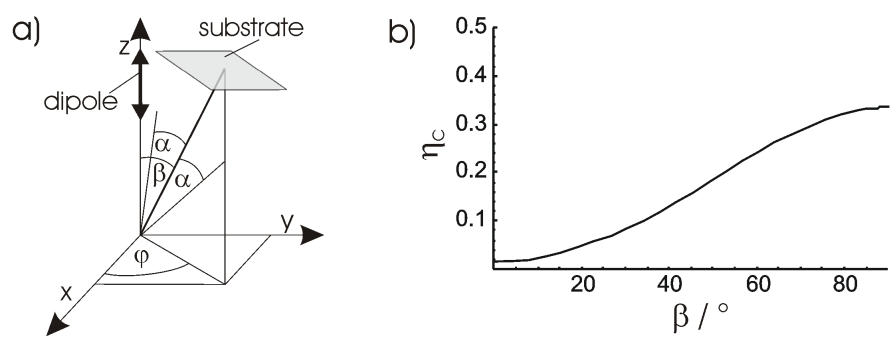

Figure S5. a) Sketch for the calculation of the illumination aperture and dipole orientation dependent collection efficiency, with $\beta$ being the angle between the emission dipole and the optical axis and $\alpha$ the maximum collection angle of the object lens (representation of the dipole in a spherical coordinate system)[11]. b) Dependency of the collection efficiency $\eta_c$ on the dipole orientation $\beta$ for an object lens aperture of $NA = 0.55$.

### VII. Estimation of the R6G surface coverage on mica from absorption spectroscopy

To determine the surface density of the molecules from an absorption measurement one needs to relate the absorbance to the surface density and the absorption cross section of the molecules. The absorption cross section of a molecule $\sigma_{abs}$ is defined as the average power $<P>$ absorbed by a molecule divided by the intensity $I$ of a plane wave incident onto the molecule [11]:

$$\sigma_{abs} = \frac{\langle P \rangle}{I} \qquad (S8)$$

In an ensemble measurement, a light beam with intensity $I$ will be attenuated by [11]

$$I(z) - I(z + dz) = -\frac{N}{V} \langle P(z) \rangle dz \qquad (S9)$$

after propagating an infinitesimal distance d$z$ through a sample with a volume concentration of molecules $N/V$. The integration of this equation yields Lambert-Beer law in its general from.[11]

Considering now a sample with $N$ molecules on a surface with the size $S$ the light beam will be attenuated by

$$I(N) - I(N + dN) = -\frac{\langle P(N) \rangle}{S} dN \qquad (S10)$$

after passing the interface. Inserting equation (S8) and integration of the equation yields

$$\ln\left(\frac{I(N)}{I_0}\right) = -\frac{N}{S}\sigma_{abs} \tag{S11}$$

with the intensity $I_0$ of the light beam incident onto the sample, the intensity $I(N)$ transmitted by the sample and the surface concentration of molecules $N/S$.

As the absorption measured with the spectrometer is here displayed as $-\log(I(N)/I_0)$ the measured signal needs to be multiplied by $\ln(10)$ to relate it to the right side of the equation above.

Fig. S6 displays absorption spectra of R6G covered (red curve) and uncovered mica substrate (black curve) performed in a double beam absorption spectrometer (UV-2101PC, Shimadzu) with blank reference. First the absorption of the R6G covered mica substrate was recorded (after spincoating of the R6G solution on the surface (front side) of the mica substrate, the rear side was cleaved off in order to remove possible R6G contamination), then after cleaving off the R6G covered surface (front side) of the mica substrate the sample was measured again in order to record the absorption of the uncovered mica substrate. The two curves were aligned in the blue/violet spectral wavelength range, in which no R6G absorption is expected.

The higher absorption (black and red curve) at shorter wavelengths is due to intrinsic absorption of the mica substrate. The oscillations (stronger at longer wavelengths) result from interference of the light beam reflections at the mica-air interfaces. After subtraction of the black curve from the red curve, the excess absorption of R6G around 530 nm becomes visible (blue curve (raw), pink curve (smoothed)), the measured signal is however close to the spectrometer accuracy (provided by the manufacturer), which results into a high experimental error. With a measured peak absorption of $0.003 \pm 0.002$ and a peak absorption cross section of R6G of $3 \pm 1 \times 10^{-16}$ cm$^2$ (peak absorption cross section of R6G in solution (unordered) is $4 \times 10^{-16}$ cm$^2$ (see ref. [14]); here incidence perpendicular to the substrate and an angle between the absorption dipole of R6G and the substrate between 45° (corresponds to unordered) and about 60° (multiplication of literature value for solution by $(\cos^2(60°)/\cos^2(45°))$ is assumed, corresponding to the orientation of cationic dyes on mica [8]), this yields a surface coverage of 0.6 - 5.8 x $10^{13}$ cm$^{-2}$.

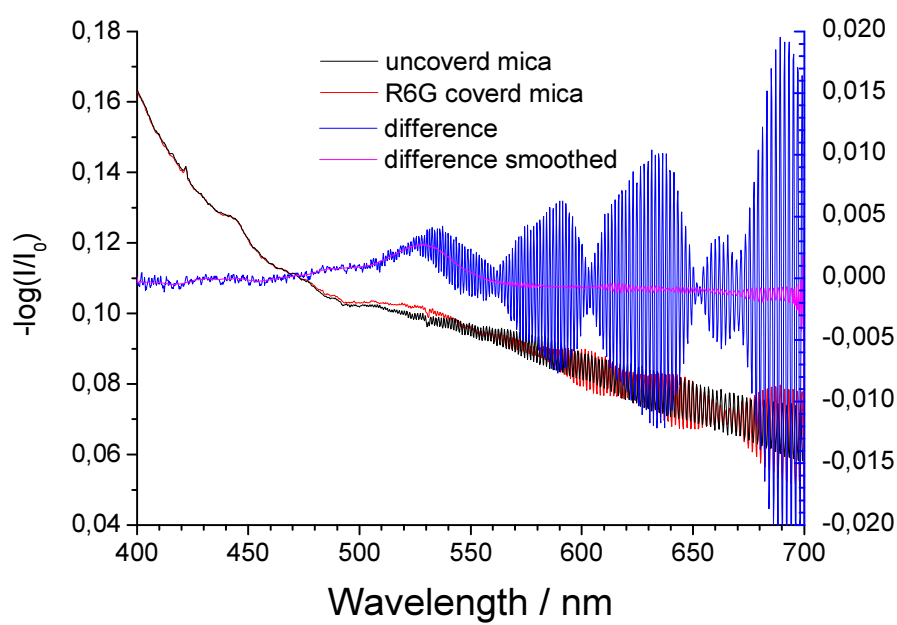

Figure S6. Left axis: Absorption spectrum of bare mica (black curve) and R6G covered mica (spincoated from 0.1 mM solution, red curve). Right axis: The difference of the absorption of R6G covered and uncovered mica (blue curve and pink curve (smoothed)) exhibits an absorption peak around 530 nm.